\documentclass[10pt]{article}
\usepackage[dvips]{graphicx}  

\begin{document}

\begin{center}
{\Large\bf Correlated  Detection of sub-mHz  Gravitational Waves by   Two Optical-Fiber Interferometers  \rule{0pt}{13pt}}\par

\bigskip
Reginald T. Cahill \\ 
{\small\it  School of Chemistry, Physics and Earth Sciences, Flinders University,
Adelaide 5001, Australia\rule{0pt}{15pt}}\\
\raisebox{+1pt}{\footnotesize E-mail: Reg.Cahill@flinders.edu.au}\par
\bigskip

Finn Stokes \\ 
{\small\it  Australian Science and Mathematics School, Flinders University,
Adelaide 5001, Australia\rule{0pt}{15pt}}\\
\raisebox{+1pt}{\footnotesize E-mail: Finn.Stokes@gmail.com}\par
\bigskip

{\small\parbox{11cm}{%
Results from two optical-fiber gravitational-wave interferometric detectors are reported.  The detector design is very small, cheap and simple to build and operate.  Using two detectors has permitted various  tests of the design principles as well as demonstrating the first  simultaneous  detection of correlated gravitational waves from  detectors spatially separated by 1.1km.  The frequency spectrum of the detected gravitational waves  is  sub-mHz with a strain spectral index $a=-1.4\pm0.1$.   As well as characterising the wave effects the detectors also show, from data collected over some 80 days in the latter part of 2007, the dominant earth rotation effect and the earth orbit effect.  The detectors operate  by  exploiting  light speed anisotropy  in   optical-fibers.   The data confirms previous observations of light speed anisotropy, earth rotation and orbit effects, and gravitational waves. \rule[0pt]{0pt}{0pt}}}\medskip
\end{center}

\setcounter{section}{0}
\setcounter{equation}{0}
\setcounter{figure}{0}
\setcounter{table}{0}

\markboth{R.T.  Cahill and F. Stokes. Correlated Detection of sub-mHz   Gravitational Waves by   Two Optical-Fiber Interferometers  }{\thepage}
\markright{R.T.  Cahill and F Stokes. Correlated Detection of sub-mHz   Gravitational Waves by   Two Optical-Fiber Interferometers }

\section{Introduction}
Results from two optical-fiber gravitational-wave interferometric detectors are reported.    Using two detectors has permitted various  tests of the design principles as well as demonstrating the first  simultaneous  detection of  correlated gravitational waves from  detectors spatially separated by 1.1km.  The frequency spectrum of the detected gravitational waves  is  sub-mHz.   As well as charactersing the wave effects the detectors also show, from data collected over some 80 days in the latter part of 2007, the dominant earth rotation effect and the earth orbit effect.  The detectors operate  by  exploiting  light speed anisotropy  in   optical-fibers.   The data confirms previous observations  \cite{MM,Miller,Illingworth,Joos, Torr,DeWitte,CahillCoax,CahillOF1,Munera} of light speed anisotropy, earth rotation and orbit effects, and gravitational waves. These observations and experimental techniques were first understood in 2002  when the Special Relativity effects and the presence of gas were used to calibrate the Michelson interferometer in gas-mode; in vacuum-mode the Michelson interferometer cannot respond to  light speed anisotropy \cite{MMCK,MMC}, as confirmed in vacuum resonant-cavity experiments, a modern version of the vacuum-mode Michelson interferometer \cite{cavities}.  The results herein come from improved versions of the prototype optical-fiber interferometer detector reported in \cite{CahillOF1}, with improved temperature stabilisation and a novel operating technique where one of  the interferometer arms is orientated with a small angular offset from the local meridian.  The detection of sub-mHz gravitational waves dates back to the pioneering work of Michelson and Morley in 1887 \cite{MM}, as discussed in \cite{Review}, and detected again by Miller  \cite{Miller} also using a gas-mode Michelson interferometer, and by Torr and Kolen \cite{Torr}, DeWitte \cite{DeWitte}  and Cahill \cite{CahillCoax}  using RF waves in coaxial cables, and by Cahill \cite{CahillOF1} and herein using an optical-fiber interferometer design, which is very much more sensitive than a gas-mode interferometer, as discussed later.

It is important to note that the repeated detection, over more than 120 years,  of the anisotropy of the speed of light  is not in conflict with the results and consequences of Special Relativity (SR), although at face value it appears to be in conflict with Einstein's 1905 postulate that the speed of light is an invariant in vacuum.  However this  contradiction is more apparent than real, for one needs to realise that the space and time coordinates  used in the standard SR Einstein formalism are {\it constructed} to make the speed of light invariant wrt those special coordinates. To achieve that observers in relative motion must then relate their space and time coordinates  by a Lorentz transformation that mixes space and time coordinates - but this is only an artifact of this formalism\footnote{Thus the detected light speed anisotropy  does not indicate a breakdown of Lorentz symmetry, contrary to the aims but  not the outcomes of \cite{cavities}.}.   Of course in the SR formalism one of the frames of reference could have always been designated as the observable   one.  That  such an ontologically real frame of reference,  only in which the speed of light is isotropic,  has been detected for over 120 years, yet ignored by mainstream physics.  The problem is in not clearly separating a very successful mathematical formalism from its predictions and experimental tests. There has been a long debate over whether the Lorentz 3-space {\it and} time interpretation or the Einstein spacetime interpretation of observed SR effects is  preferable or indeed even experimentally distinguishable.  

What has been discovered in recent years is that a dynamical structured 3-space exists, so confirming the Lorentz interpretation of SR, and with fundamental implications for physics - for  physics failed to notice the existence of the main {\it constituent} defining the universe, namely a dynamical 3-space, with quantum matter and EM radiation playing a minor role.  This dynamical 3-space provides an explanation for the success of the SR Einstein formalism.   It also provides a new account of gravity, which turns out to be a quantum effect \cite{Schrod}, and of cosmology  \cite{Review,Book,Hubble, QC}, doing away with the need for dark matter and dark energy.

\begin{figure}[t]
\vspace{27mm}
\hspace{15mm}
\setlength{\unitlength}{1.9mm}
\hspace{0mm}\begin{picture}(0,0)
\thicklines
\put(4,0.0){\line(1,0){10}}
\put(4,2){\line(1,0){10}}
\put(4,0){\line(0,1){2}}
\put(14,0){\line(0,1){2}}
\put(6,5.5){\bf He-Ne}
\put(6,+3){\bf laser}

\put(19,1.25){\line(-2,-1){4}}
\put(18,1.25){\line(1,0){10}}
\put(18,1.15){\line(1,0){10}}
\put(18,1.05){\line(1,0){10}}
\put(18,0.95){\line(1,0){10}}
\put(18,0.85){\line(1,0){10}}
\put(18,0.75){\line(1,0){10}}
\put(18,0.75){\line(0,1){0.5}}
\put(28,0.75){\line(0,1){0.5}}
\put(19.5,5.5){\bf 2x2 beam-}
\put(19.5,3.1){\bf splitter}
\put(14,1.0){\line(1,0){4}}
\put(14,1.0){\vector(1,0){3}}

\put(4,-5){\line(1,0){4}}
\put(4,-3){\line(1,0){4}}
\put(4,-5){\line(0,1){2}}
\put(8,-5){\line(0,1){2}}
\put(4,-8){\bf photodiode}
\put(4,-10.5){\bf detector}

\put(-4,-5){\line(1,0){4}}
\put(-4,-3){\line(1,0){4}}
\put(-4,-5){\line(0,1){2}}
\put(0,-5){\line(0,1){2}}
\put(-5,-8){\bf data}
\put(-5,-10.5){\bf logger}
\put(0,-4){\line(1,0){4}}

\put(19,-3.75){\line(-2,-1){4}}
\put(18,-3.75){\line(1,0){10}}
\put(18,-3.85){\line(1,0){10}}
\put(18,-3.95){\line(1,0){10}}
\put(18,-4.05){\line(1,0){10}}
\put(18,-4.15){\line(1,0){10}}
\put(18,-4.25){\line(1,0){10}}

\put(18,-4.25){\line(0,1){0.5}}
\put(28,-4.25){\line(0,1){0.5}}
\put(19.5,-8.0){\bf 2x2 beam-}
\put(19.5,-10.5){\bf joiner}
\put(8,-4){\line(1,0){10}}
\put(16,-4){\vector(-1,0){3}}
\put(28,4.4){\oval(40,6)[r,b]}
\put(50.5,4.2){\oval(5,40)[l,t]}
\put(50,-1){\oval(5,50.4)[t,r]}
\put(50.3,-0.9){\oval(4.5,4.5)[r,b]}
\put(50.5,-0.6){\oval(4.,5.)[l,b]}
\put(50.5,-0.6){\oval(4.,48)[l,t]}
\put(50.5,-2.6){\oval(3.,52)[r,t]}
\put(28,-1.65){\oval(48,4)[b,r]}

\put(28,1.4){\vector(1,0){5}}
\put(28,0.7){\vector(1,0){7}}
\put(28,-2.0){\oval(90,5.5)[r,t]}
\put(49.0,-1.85){\oval(48.0,3)[b,r]}
\put(50.0,-1.40){\oval(5,3.9)[l,b]}
\put(70.0,-1.7){\oval(45,4)[l,t]}
\put(69.5,-2.2){\oval(8.0,5)[t,r]}
\put(28,-1.8){\oval(91,5)[b,r]}

\put(38,-3.7){\vector(-1,0){5}}
\put(38,-4.35){\vector(-1,0){7}}
\put(48.3,-0.3){\line(1,0){0.5}}
\put(48.3,-1.3){\line(1,0){0.5}}
\put(48.3,-2.3){\line(1,0){0.5}}
\put(48.3,-2.3){\line(0,1){2}}
\put(48.8,-2.3){\line(0,1){2}}
\put(40,-8.0){\bf FC mating} 
\put(40,-10.5){\bf sleeves}
\put(37,10){\bf ARM 1}
\put(47.3,-0.3){\line(1,0){0.5}}
\put(47.3,-1.3){\line(1,0){0.5}}
\put(47.3,-2.3){\line(1,0){0.5}}
\put(47.3,-2.3){\line(0,1){2}}
\put(47.8,-2.3){\line(0,1){2}}
\put(58,2.0){\bf ARM 2}

\put(52.0,-5.5){\vector(1,0){19}}
\put(58.0,-5.5){\vector(-1,0){8}}
\put(56,-9.0){\bf 100mm}

\put(70,-2.0){\bf o}
\put(49.5,-2.0){\bf o}
\put(49.5,21.0){\bf o}


\end{picture}
\vspace{20mm}
\caption{  {Schematic layout of the interferometric optical-fiber  light-speed anisotropy/gravitational wave detector.  Actual detector   is shown  in Fig.\ref{fig:detphoto}.  Coherent 633nm light from the a He-Ne Laser is split into two  lengths of single-mode  polarisation preserving fibers by the 2x2 beam splitter. The two fibers take different directions, ARM1 and ARM2, after which the light is recombined in the 2x2 beam joiner,  in which the phase differences lead to interference effects  that are indicated by the outgoing light intensity, which is measured in the photodiode detector/amplifier (Thorlabs PDA36A or PDA36A-EC), and then recorded in the data logger.  In the actual layout the fibers make two loops in each arm, but with  excess lengths wound around one arm (not shown) - to reduce effective fiber lengths so as to reduce sensitivity. The length of one straight section is 100mm, which is the center to center spacing of the plastic turners, having diameter = 52mm, see Fig.\ref{fig:detphoto}.   The relative travel times, and hence the output light intensity, are affected by the varying speed and direction of the flowing 3-space, by affecting differentially the speed of the light, and hence the net phase difference between the two arms.   }}

 \label{fig:schematic}
\end{figure}
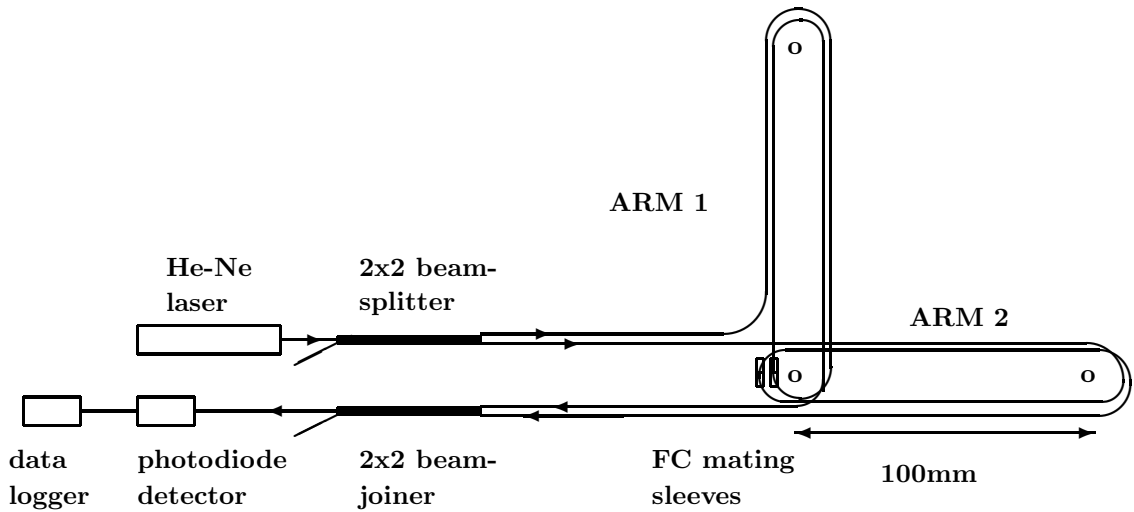

\begin{figure}[t]
\vspace{0mm}
\hspace{30mm}\includegraphics[height=100mm,width=120mm,angle=90,keepaspectratio=false]{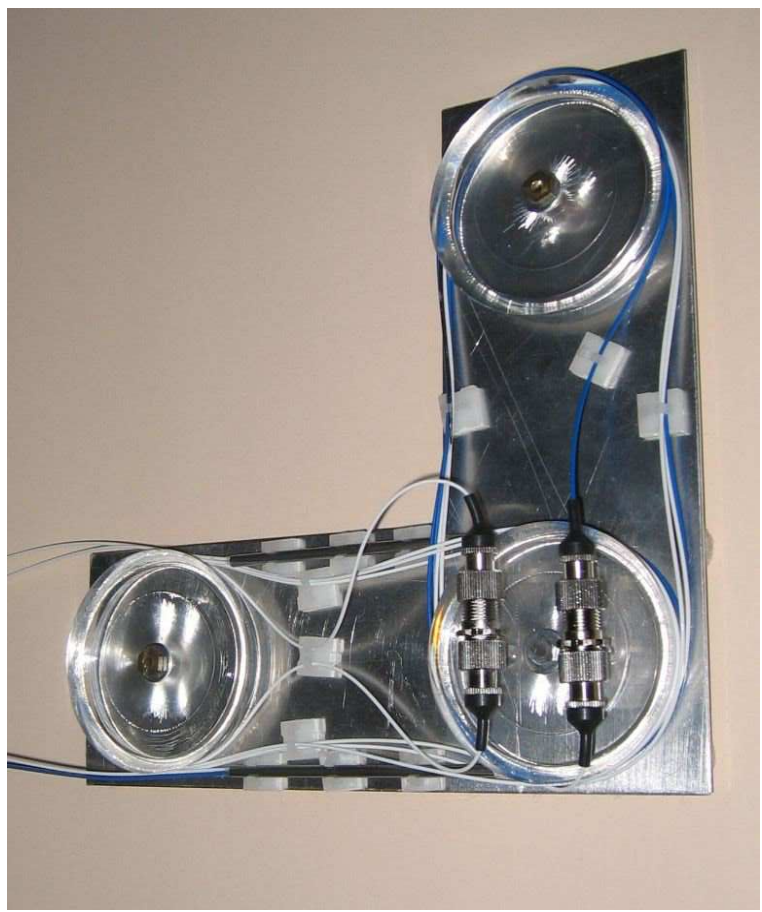}
\vspace{0mm}
\caption{{Photograph of a detector showing the optical fibers forming the two orthogonal arms. See Fig.\ref{fig:schematic} for the schematic layout.  The 2x2 beam splitter and joiner  (Thorlabs FC632-50B-FC) are the two  small stainless steel cylindrical tubes. The two FC to FC mating sleeves (Thorlabs ADAFC1)  are physically adjacent.   The overall dimensions of the metal base plate are 160mm$\times$160mm. The 2$\times$2 splitter and joiner each have two input and two output fibers, with one not used.  Arm 2 is folded over the splitter and joiner, compared to the schematic layout. The interferometer shown costs approximately \$400.}}
  \label{fig:detphoto}
\end{figure}

\begin{figure}

\vspace{-10mm}

\hspace{35mm}\includegraphics[scale=0.40]{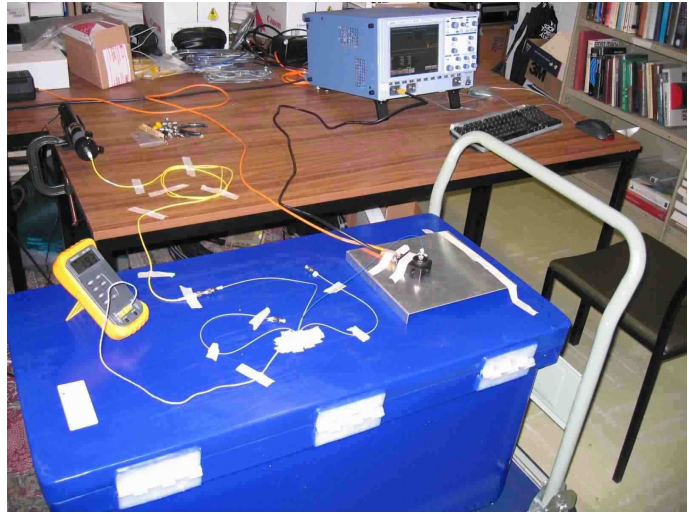}

\vspace{10mm}
\hspace{35mm}\includegraphics[scale=0.4]{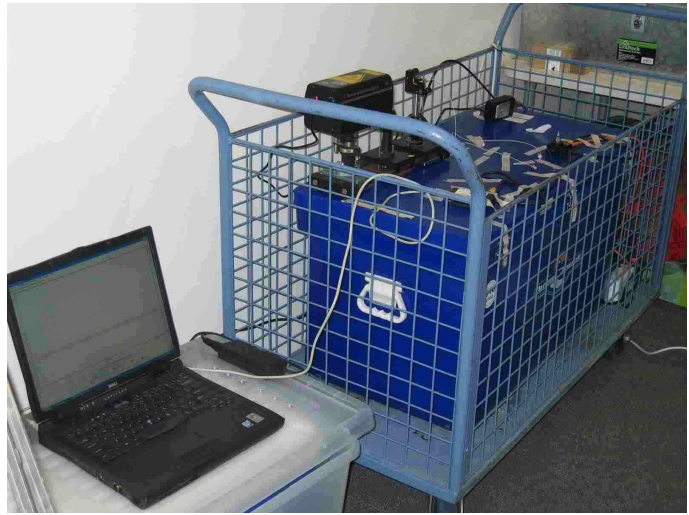}
\vspace{0mm}
\caption{{(a) Detector  1  (D1) is located  inside a sealed air-filled bucket inside an insulated container (blue)  containing some 90kg of water for temperature stabilisation. This detector, in the School of Chemistry, Physics and Earth Sciences, had an orientation of $5^0$ ant-iclockwise to the local meridian. Cylindrical He-Ne laser (Melles-Griot 0.5mW  633nm 05-LLR-811-230) is located on LHS of bench, while data logger is on RHS.  Photodiode detector/pre-amplifier is located atop aluminium plate.  (b) Detector 2  (D2) was located 1.1km North of D1 in the Australian Science and Mathematics School.  This detector had an orientation of  $11^0$ anti-clockwise to the local meridian.   The data was logged on a PC running  a  PoScope USB DSO  (PoLabs http://www.poscope.com).
 \label{fig:detector1}}}
\end{figure}

\begin{figure}
\vspace{-10mm}
\hspace{15mm}\includegraphics[scale=0.60]{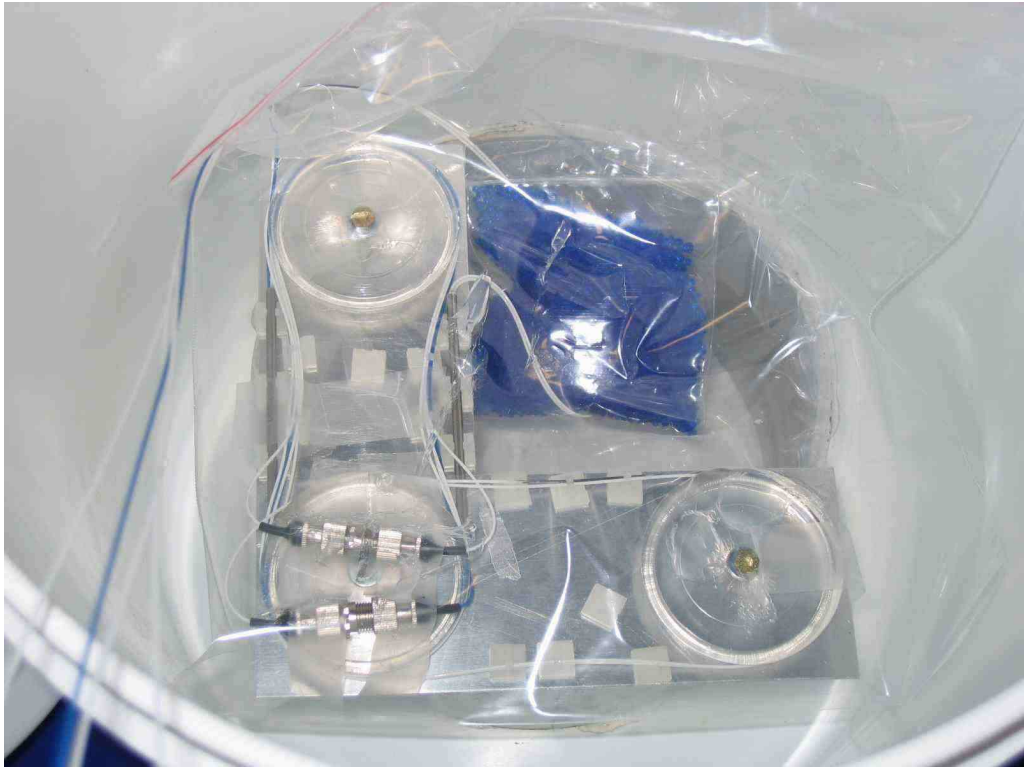}

\vspace{5mm}\hspace{40mm}\includegraphics[scale=0.70]{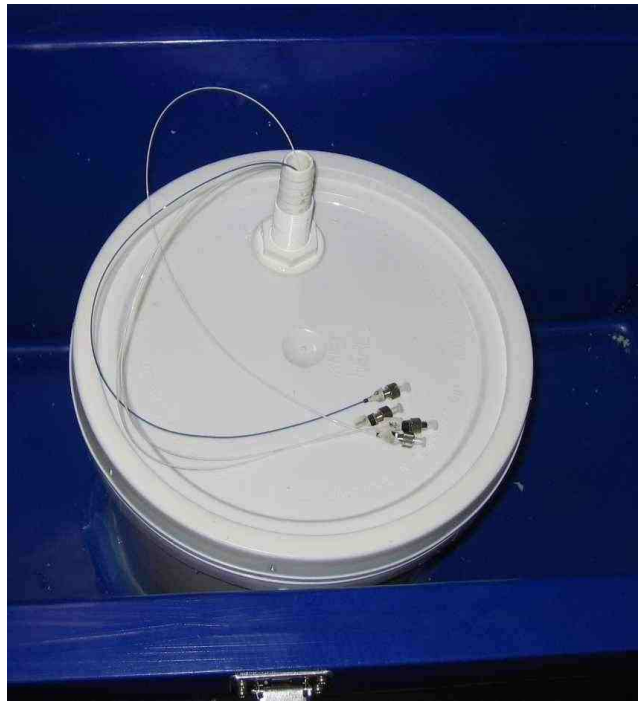}

\vspace{7mm}
\caption{{(a) Detectors are horizontally located inside an air-filled bucket. The plastic bag reduces even further any air  movements, and thus temperature differentials. The blue crystals are silica gel to reduce moisture.  (b) Bucket located inside and attached to bottom of the insulated container prior to adding water to the container.
  \label{fig:inside}}}
\end{figure}

\begin{figure}
\vspace{-2mm}
\hspace{20mm}\includegraphics[scale=0.60]{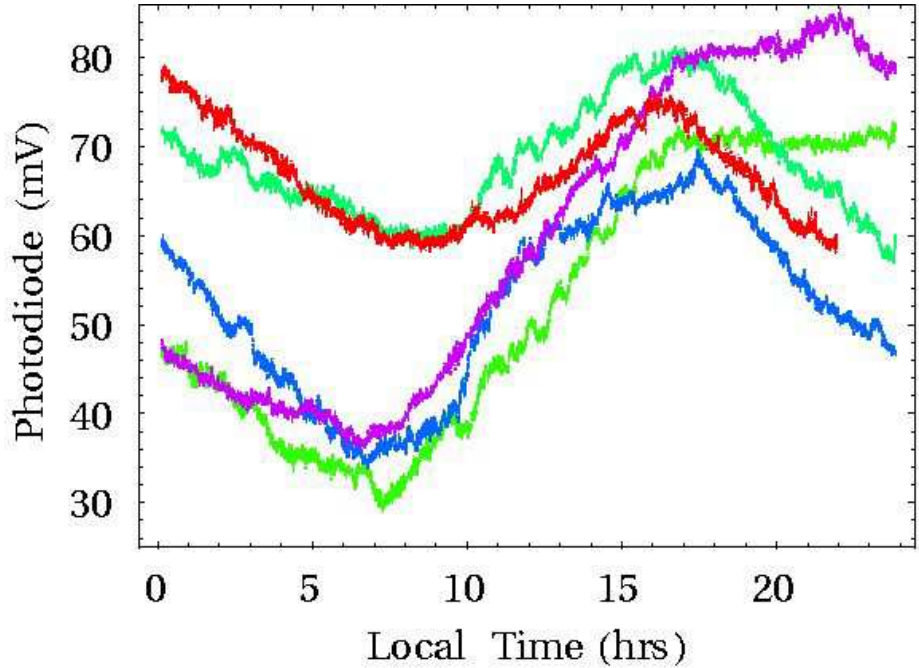}
\vspace{0mm}
\caption{{D1 photodiode output voltage  data (mV), recorded every 5 secs,  from 5 successive days,  starting September 22, 2007, plotted against local Adelaide time (UT= local time + 9.5hrs).  Day sequence is indicated by increasing hue.  Dominant minima and maxima is earth rotation effect. Fluctuations from day to day are evident as are fluctuations during each day - these are caused by wave effects in the flowing space.  Changes in RA cause changes in timing of min/max, while changes in magnitude are caused by changes in declination and/or speed.  Blurring effect is caused by laser noise. Same data   is plotted sequentially in Fig.\ref{fig:DataSeq}a.
  \label{fig:5days}}}
\end{figure}

\begin{figure}
\vspace{-5mm}
\hspace{40mm}\includegraphics[height=65mm,width=70mm,keepaspectratio=false]{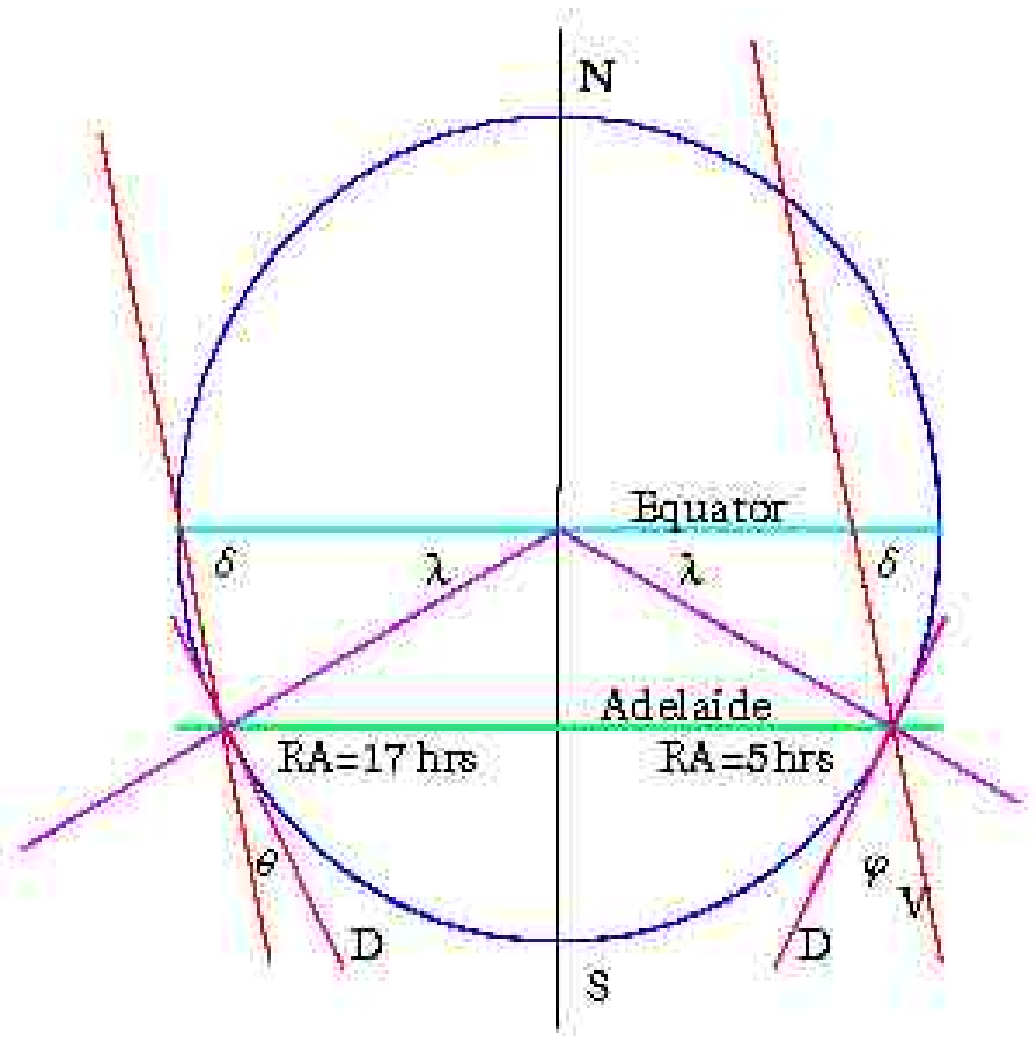}
\vspace{-3mm}
\caption{{Schematic of earth and spatial flow at approximate local sidereal times (RA) of 5hrs  and 17hrs.  The detector arms, D,  of D1 and D2 are operated at  small offset angles from the local meridian.  The long straight lines indicate the spatial flow velocity vector, with declination $\delta$.   The large earth-rotation induced minima/maxima are caused by the inclination angle varying from a maximum  $\phi$ to a minimum  $\theta$, respectively.  Wave effects are changes in the velocity vector.  
  \label{fig:earth}}}
\end{figure}

\section{Dynamical 3-Space and Gravitational Waves}
Light-speed anisotropy experiments have revealed that a dynamical 3-space exists, with the speed of light being $c$, in vacuum, only wrt to this space: observers in motion `through' this 3-space detect that the speed of light is in general different from $c$, and is different in different directions\footnote{Many failed experiments supposedly designed to detect this anisotropy can be shown to have design flaws.}.  The dynamical equations for this 3-space are now known and involve a velocity field ${\bf v}({\bf r},t)$, but where only relative velocities are observable locally  - the coordinates ${\bf r}$ are relative to a non-physical mathematical embedding space. These dynamical equations involve Newton's gravitational constant $G$ and the fine structure constant $\alpha$.  The discovery of this dynamical 3-space then required a generalisation of the Maxwell, Schr\"{o}dinger and Dirac equations.     The wave effects already detected correspond to fluctuations in the 3-space velocity field ${\bf v}({\bf r},t)$, so they are really 3-space turbulence or wave effects.  However they are better known, if somewhat inappropriately, as `gravitational waves' or `ripples' in `spacetime'.  Because the 3-space dynamics gives a deeper understanding of the spacetime formalism we now know that the metric of the induced spacetime, merely a mathematical construct having no ontological significance, is related to ${\bf v}({\bf r},t)$ according to \cite{Review,Book,QC}
\begin{equation}
ds^2=dt^2 -(d{\bf r}-{\bf v}({\bf r},t)dt)^2/c^2
=g_{\mu\nu}dx^{\mu}dx^\nu
\label{eqn:Eqn1}\end{equation}
The gravitational acceleration of matter, and of the structural patterns  characterising  the 3-space, are given by \cite{Review,Schrod}
\begin{equation}
{\bf g}=\frac{\partial {\bf v}}{\partial t}+({\bf v}.\nabla ){\bf v}
\label{eqn:acceln}
\end{equation}
and so fluctuations in  ${\bf v}({\bf r},t)$ may or may not manifest as a gravitational force.
The general characteristics of   ${\bf v}({\bf r},t)$ are now known following the detailed analysis  of the experiments noted above, namely its average speed,  and removing the earth orbit effect, is  some 420$ \pm $30km/s, from direction RA= $5.5 \pm 2^{hr}$, Dec=$70 \pm 10^o$S -  the center point of the Miller data in Fig.\ref{fig:FT}b, together with large wave/turbulence effects.  The magnitude of this turbulence depends on the timing resolution of each particular experiment, and here we characterise them at sub-mHz frequencies, showing  that the  fluctuations are very large, as also seen in \cite{CahillCoax}.  

\begin{figure}
\vspace{-4mm}
\hspace{0mm}\includegraphics[scale=0.78]{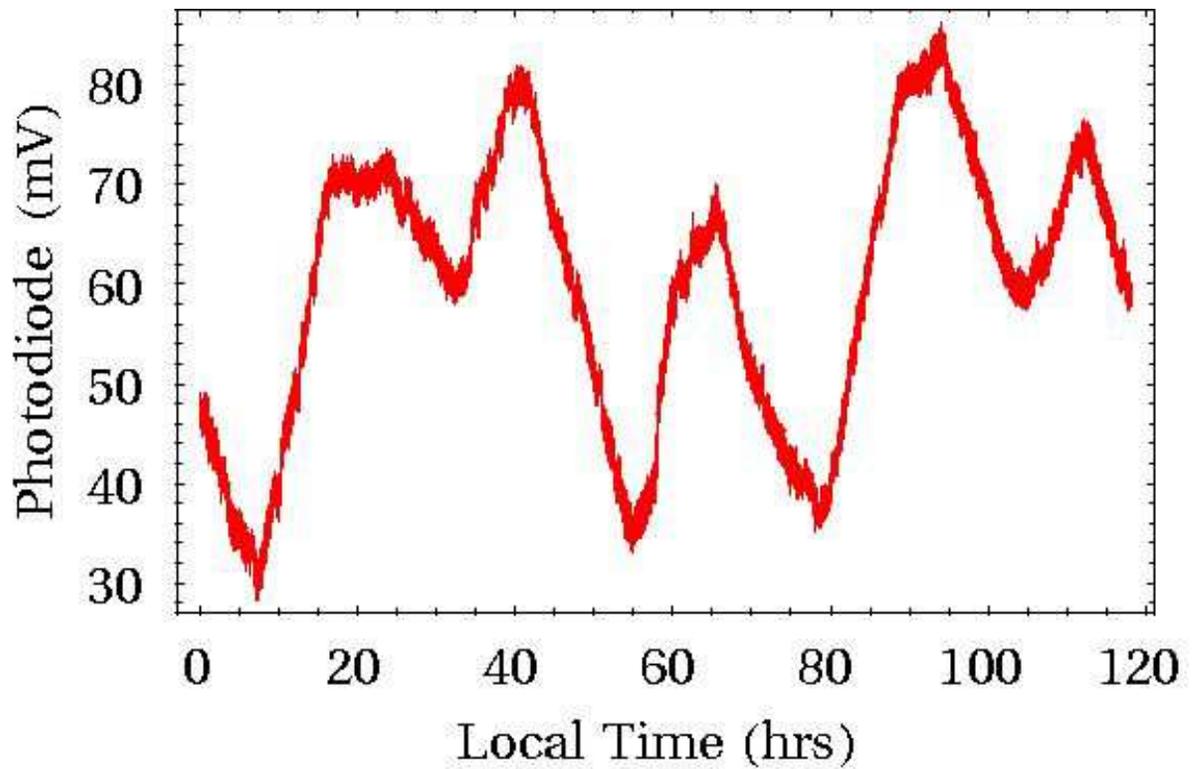}

\vspace{3mm}
\hspace{-1mm}\includegraphics[scale=0.80]{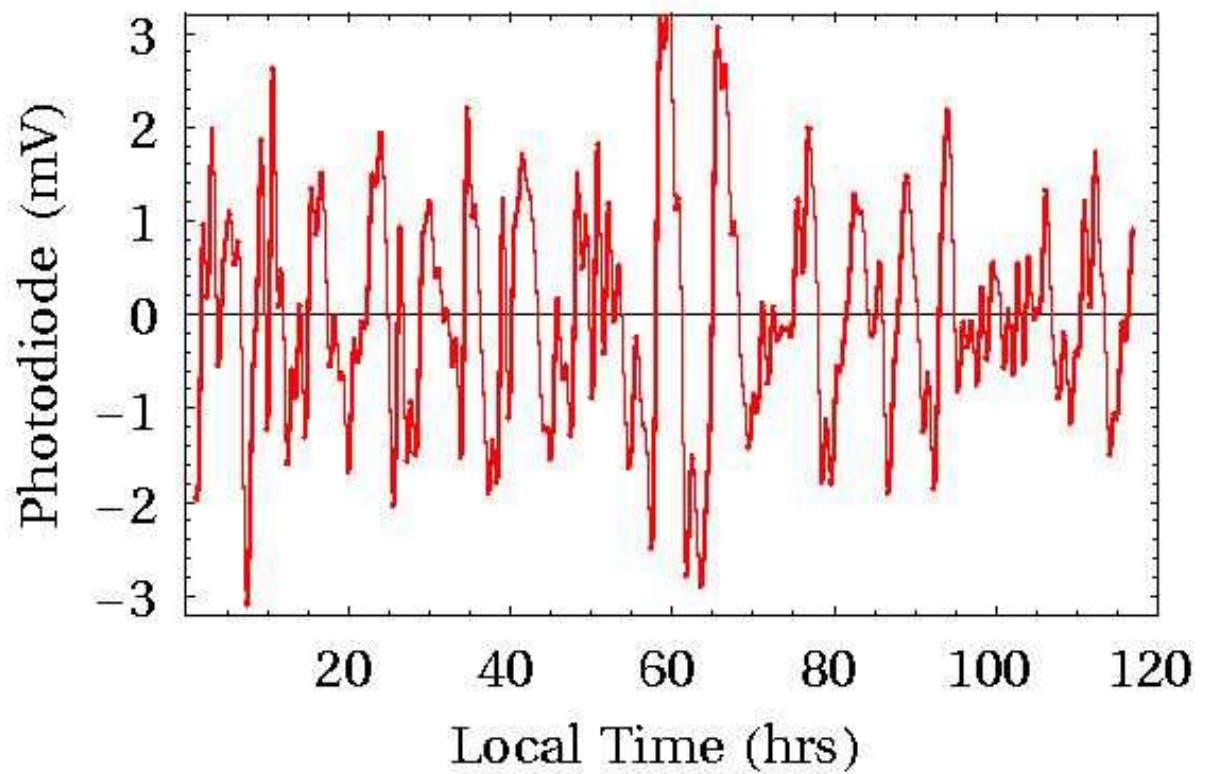}
\caption{{(a)  Plot of 5 days of data from Fig.\ref{fig:5days} shown sequentially.  The Fourier Transform of this data is shown in Fig.\ref{fig:FT}a.  (b) Plot shows  data after filtering out earth-rotation frequencies (f $<$ 0.025mHz) and laser noise frequencies (f $>$ 0.25mHz, Log$_{10}$[0.25]=-0.6).  This shows wave/turbulence effects.  Note that the magnitude of the wave component of the signal is some 10$\%$ of full signal in this bandwidth. \label{fig:DataSeq}}}
\end{figure}

\section{Gravitational Wave Detectors}
To measure  ${\bf v}({\bf r},t)$ has been difficult until now. The early experiments used gas-mode Michelson interferometers, which involved the visual observation of small fringe shifts as the relatively large devices were rotated.  The RF coaxial cable experiments had the advantage of permitting electronic recording of the RF travel times, over 500m \cite{Torr} and 1.5km \cite{DeWitte}, by means of two or more atomic clocks, although the experiment reported in \cite{CahillCoax} used a novel technique that enable the coaxial cable length to be reduced to laboratory size\footnote{The calibration of this technique is at present not well understood in view of recent discoveries concerning the Fresnel drag effect in optical fibers.}.  The new  optical-fiber detector design herein has the  advantage of electronic recording as well as high precision because the travel time differences in the two orthogonal fibers employ light interference effects, but with the interference effects taking place in an optical fiber  beam-joiner, and so no optical projection problems arise.  The device is very  small, very cheap and easily assembled from readily available opto-electronic components. The schematic  layout of the detector is given in Fig.\ref{fig:schematic}, with a detailed description in the figure caption.  The detector relies on the phenomenon where the 3-space velocity ${\bf v}({\bf r},t)$ affects differently the light travel times in the  optical fibers, depending on the projection of ${\bf v}({\bf r},t)$ along the fiber directions. The differences in the light travel times are measured by means of the interference effects in the beam joiner. The difference in travel times is given by  
\begin{equation}
\Delta t=k^2\frac{Lv_P^2}{c^3}\cos\bigl(2\theta\bigr)
\label{eqn:times}\end{equation}   
where 
$$k^2=\frac{(n^2-1)(2-n^2)}{n}$$
is the instrument calibration constant, obtained by taking account of the three key effects: (i) the different light paths, (ii) Lorentz contraction of the fibers, an effect depending on the angle of the fibers to the flow velocity, and (iii) the refractive index effect, including the Fresnel drag effect.  Only if $n\neq 1$ is there a net effect, otherwise when $n=1$ the various effects actually cancel. So in this regard the Michelson interferometer has a serious design flaw. This problem has been overcome by using optical fibers. Here $n=1.462$ at $633$nm is the effective refractive index of the single-mode optical fibers (Fibercore SM600, temperature coefficient $5\times10^{-2}$ fs/mm/C).  Here $L\approx 200mm$ is the average effective length of the two arms, and $v_P({\bf r},t)$ is the projection of ${\bf v}({\bf r},t)$ onto the plane of the detector, and the angle $\theta$ is that of the projected velocity onto the arm.   

The reality of the Lorentz contraction effect is experimentally confirmed by comparing the 2nd order in $v/c$ Michelson gas-mode interferometer data, which requires account be taken of the contraction effect,  with that from the 1st order in $v/c$  RF coaxial cable travel time experiments, as in DeWitte \cite{DeWitte}, which does not require that the contraction effect be taken into account, to give comparable values for $v$.
 
 For gas-mode Michelson interferometers  $k^2\approx n^2-1$, because then  $n\approx 1^+$  is the refractive index of a gas.  Operating in air, as for Michelson and Morley and for Miller, n=1.00029, so that $k^2=0.00058$, which in particular means that the Michelson-Morley interferometer was nearly 2000 times less sensitive than assumed by Michelson, who used Newtonian physics to calibrate the interferometer - that analysis gives $k^2=n^3 \approx 1$. Consequently the small fringe shifts observed  by Michelson and Morley actually correspond to a light speed anisotropy of some 400 km/s, that is, the earth has that speed relative to the local dynamical 3-space. The dependence of $k$ on $n$ has been checked \cite{MMCK, Book} by comparing the  air gas-mode data against data from the He gas-mode operated interferometers of Illingworth \cite{Illingworth} and Joos \cite{Joos}. 

The above analysis also has important implications for long-baseline terrestrial vacuum-mode Michelson interferometer gravitational wave detectors - they have a fundamental design flaw and will not be able to detect gravitational waves.

The interferometer operates by detecting changes in the travel time difference between the two arms,  as given by (\ref{eqn:times}). The cycle-averaged light intensity emerging from the beam joiner is given by
\begin{eqnarray}
I(t)&\propto&\left(Re({\bf E}_1+{\bf E}_2e^{i\omega( \tau +\Delta t})\right)^2 \nonumber \\
&=&2|{\bf E}|^2\cos\left(\frac{\omega(\tau+\Delta t)}{2}\right)^2   \nonumber \\
&\approx& a+b\Delta t
\label{eqn:intensity}\end{eqnarray}
Here ${\bf E}_i$ are the electric field amplitudes  and have the same value as the fiber splitter/joiner are 50\%-50\% types,  and having the same direction because polarisation preserving fibers are used,   $\omega$ is the light angular frequency and $\tau$ is a travel time difference caused by the light travel times not being identical, even when $\Delta t=0$, mainly because the various splitter/joiner fibers will not be identical in length. The last expression follows because $\Delta t$ is small, and so the detector operates, hopefully,  in a linear regime, in general, unless $\tau$ has  a value equal to modulo($T$), where $T$ is the light period.  The main temperature effect in the detector, so long as a  temperature uniformity is maintained, is that $\tau$ will be temperature dependent. The temperature  coefficient for the optical fibers gives an effective fractional fringe shift error of $\Delta\tau/T=3\times10^{-2}$/mm/C, for each mm of length difference.  The photodiode detector output voltage  $V(t)$  is proportional to $I(t)$, and so finally linearly related to $\Delta t$.  The detector calibration constants $a$ and $b$ depend on $k$, $\tau$ and the laser intensity  and are unknown at present. 

\section{Data Analysis}
The data is described in detail in the figure captions.

\noindent Fig.\ref{fig:5days} shows 5 typical days of data exhibiting  the earth-rotation effect, and also fluctuations during each day and from day to day, revealing  dynamical 3-space turbulence - essentially the long-sort-for gravitational waves. It is now known that these gravitational waves  were first detected in the Michelson-Morley 1887 experiment \cite{Review}, but only because their interferometer was operated in gas-mode.  Fig.\ref{fig:FT}a shows the frequency spectrum for this data. 

\noindent Fig.\ref{fig:DataSeq}b shows the gravitational waves after removing frequencies near the earth-rotation frequency. As discussed later these gravitational waves are predominately sub-mHz.

\begin{figure}
\vspace{-3mm}

\hspace{0mm}\includegraphics[scale=0.50]{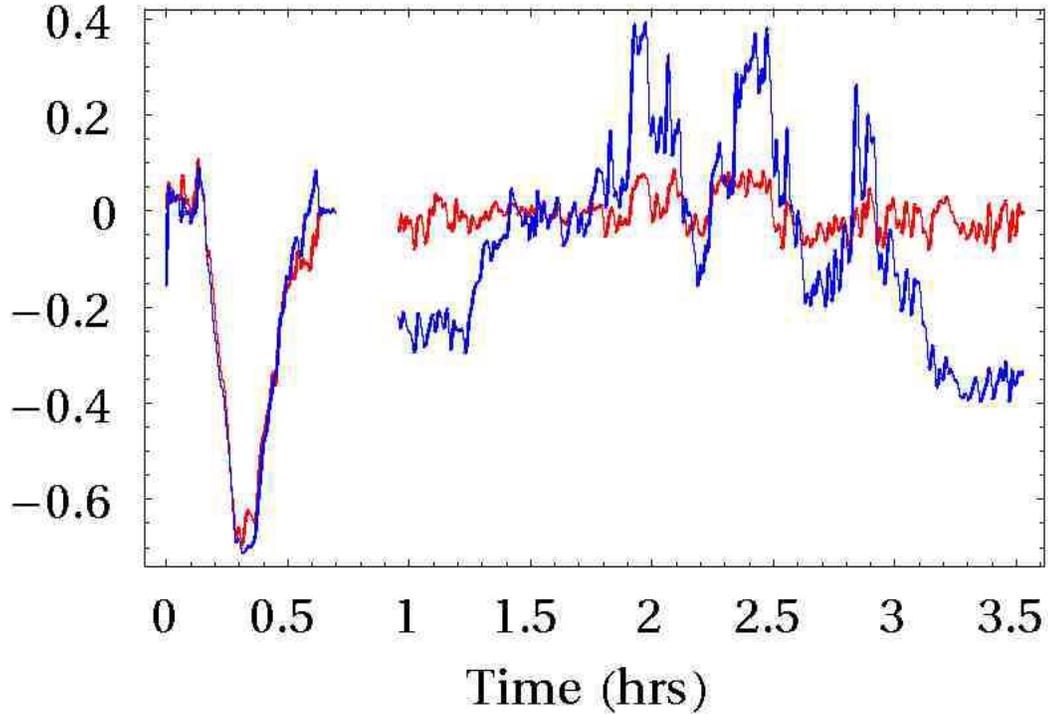}
\vspace{-0mm}
\caption{{ Two tests of the detectors.  (a) The left plot shows data from D1 and D2 when co-located, parallel and operating from the same laser.   The data from one has been rescaled to match the data from the other, as they have different calibrations.  Both detectors  show a simultaneous gravitational wave pulse of duration $\approx$ 0.5hrs.  (b) The  right plot shows data from D2 (blue) and  from a direct feed of the common laser source to the photodiode detector of D1 (red), i.e bypassing the D1 interferometer.  This data has been rescaled so that high frequency components have the same magnitude, to compensate for different feed amplitudes. The laser-only signal  (red) shows the amplitude and frequency noise level  of the laser .  The signal from D2 (blue) shows the same noise effects, but with additional larger variations - these are wave effects detected by  D2 operating in interferometer mode. This data shows that the laser noise is dominant above approximately 1mHz.}
  \label{fig:Correlations1}}
\end{figure}

\begin{figure}
\vspace{-20mm}
\hspace{30mm}\includegraphics[scale=0.45]{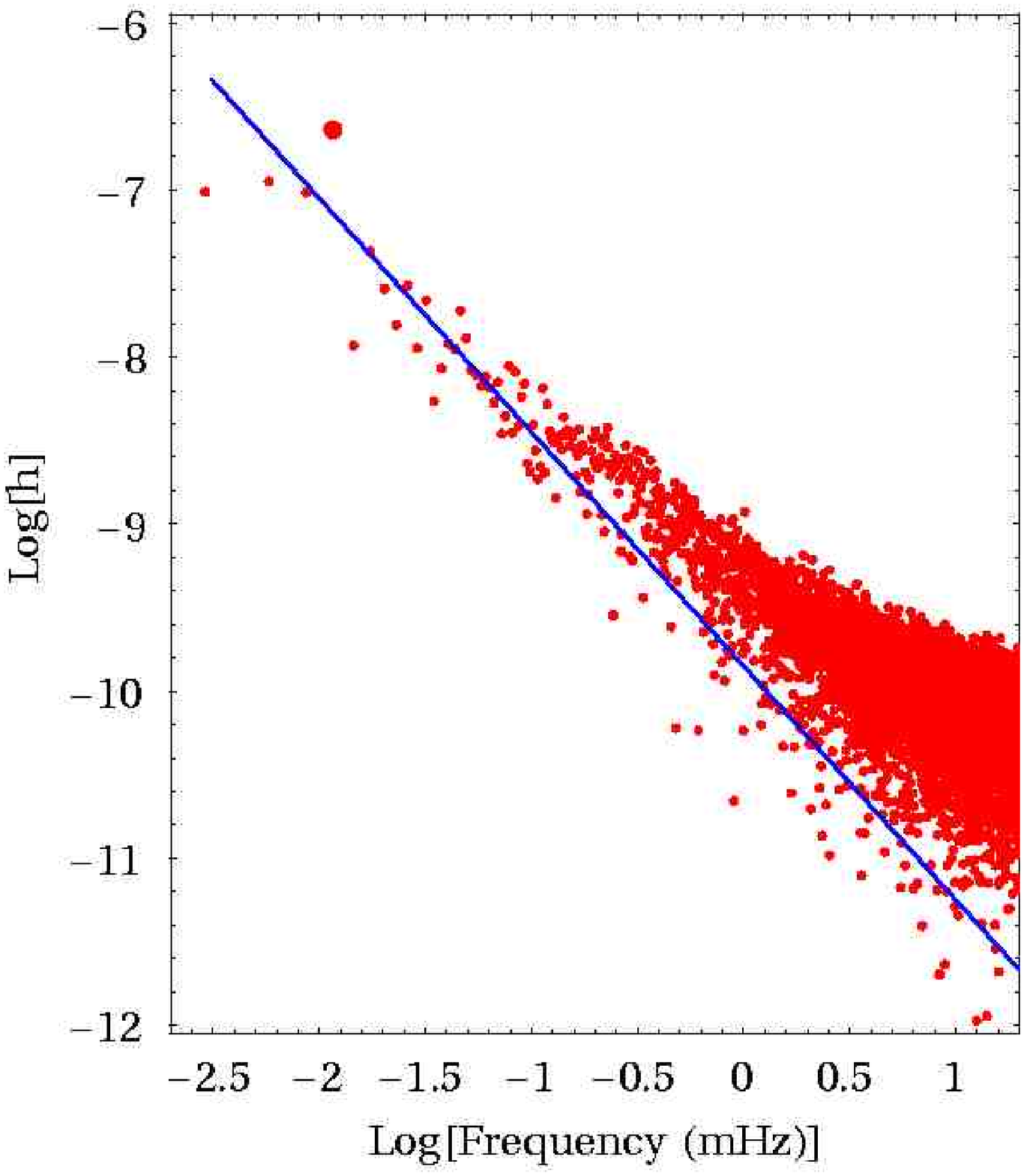}

\hspace{45mm}\includegraphics[scale=0.25]{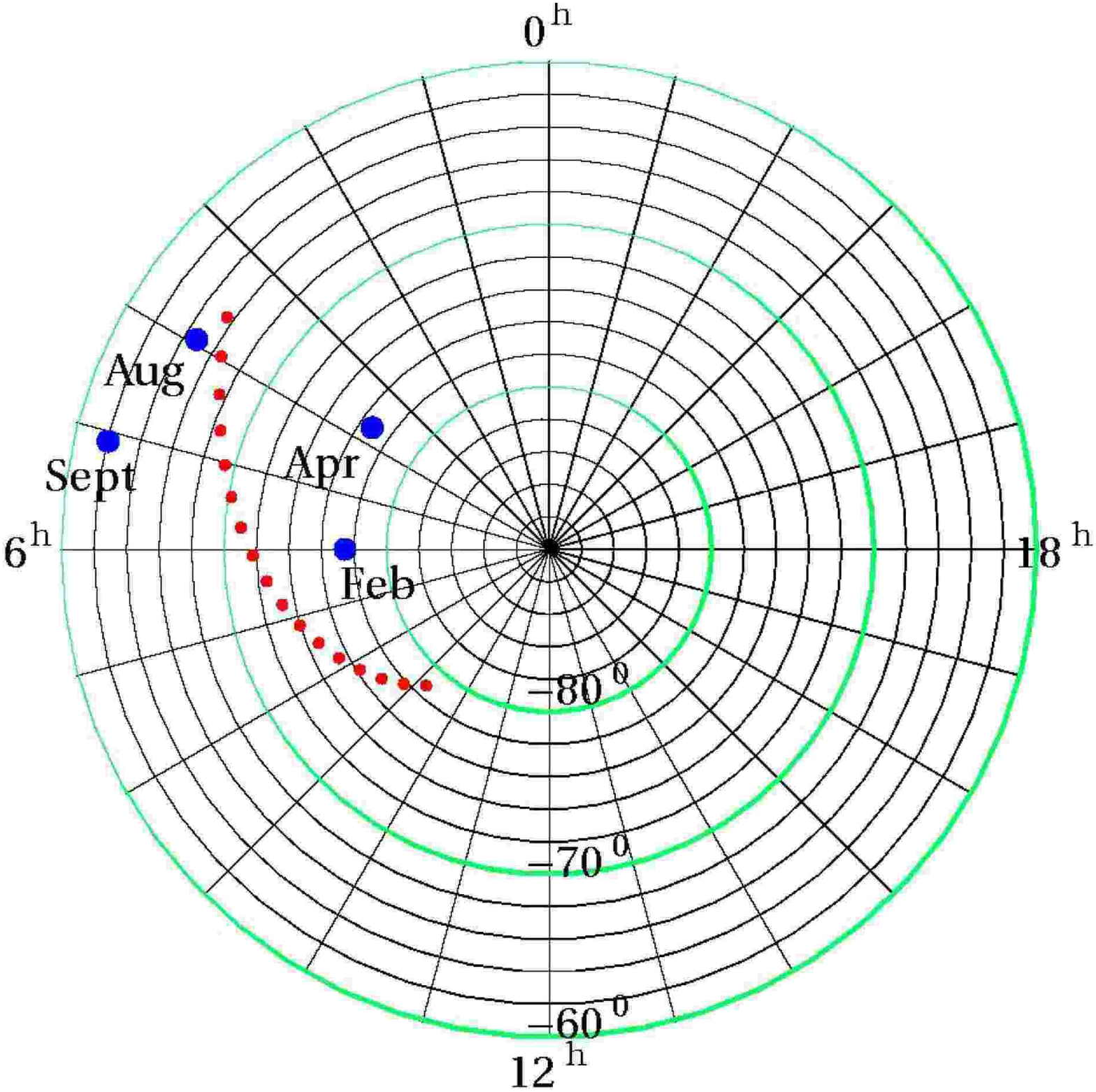}
\vspace{0mm}
\caption{{ (a) Log-Log plot of the frequency spectrum $|h(f)|$ of the data from the five days shown in Fig.\ref{fig:DataSeq}a.  $h(f)$ is the strain $v^2/2c^2$ at frequency $f$, normalised to $v=400$km/s at the 24hr frequency.  The largest component (large red point) is the 24hr earth rotation frequency. The straight line (blue) is a trend line that suggests that the signal has two components - one indicated by the trend line having the form $|h(f)|\propto f^a$ with strain spectral index $a=-1.4\pm0.1$, while the second component, evident above 1mHz,  is noise from the laser source, as also indicated by the data in Fig.\ref{fig:Correlations1}. (b) Southern celestial sphere with RA and Dec shown. The 4 blue points show the results from Miller \cite{Miller} for four months in 1925/1926.  The sequence of red points show the daily averaged RA and Dec as determined from the data herein for every 5 days.  The 2007 data shows a direction that moves closer to the south celestial pole in late December than would be indicated by the Miller data.  The new results differ by  10$^0$ to 13$^0$ from the corresponding Miller data points (the plot exaggerates angles). The wave effects cause the actual direction to fluctuate from day to day and during each day.}
  \label{fig:FT}}
\end{figure}

\noindent Fig.\ref{fig:Correlations1} reports one of a number of key experimental tests of the detector principles. These show the two detector responses when (a) operating from the same laser source, and  (b) with only D2 operating in interferometer mode.  These reveal the noise effects coming from the laser in comparison with the interferometer signal strength.  This gives a guide to the S/N ratio of these detectors. 

\begin{figure}
\vspace{-3mm}
\hspace{20mm}\includegraphics[scale=0.60]{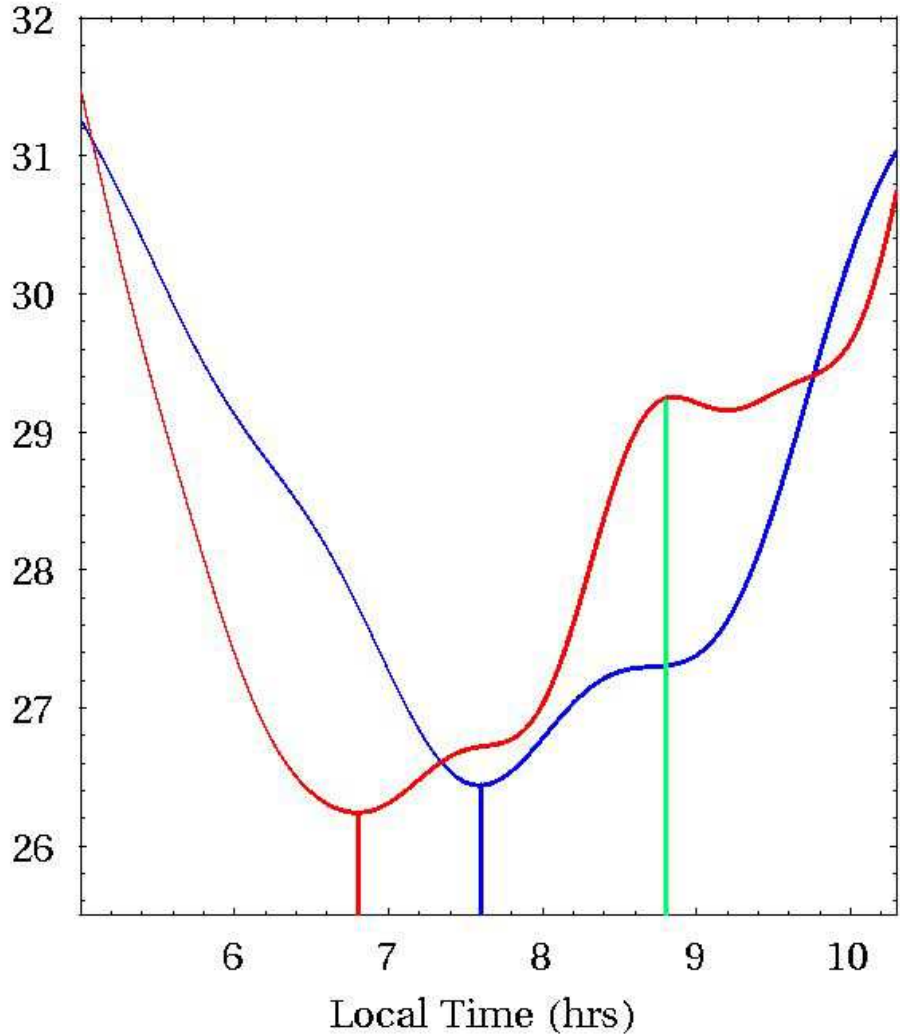}
\vspace{-3mm}
\caption{{ Photodiode data (mV) on October 4, 2007,  from  detectors D1 (red plot) and D2  (blue plot) operating simultaneously with D2 located 1.1km due north of D1.   A low-pass  FFT filter (f $\leq$ 0.25mHz, Log$_{10}$[f(mHz)] $\leq$ -0.6) was used to remove  laser noise.  D1 arm is aligned $5^0$ anti-clockwise from local meridian, while D2 is aligned $11^0$ anti-clockwise from local meridian. The alignment offset between D1 and D2 causes the dominant earth-rotation induced minima to occur at {\it different} times, with that of D2  at t = 7.6hrs {\it delayed} by 0.8hrs relative to D1 at t = 6.8hrs, as expected from Figs.\ref{fig:MaxMin}b and \ref{fig:MidMaxMin}b for  Dec = 77$^0$.  This is a fundamental test of the detection theory and of the phenomena.  As well the data shows a {\it simultaneous}  sub-mHz gravitational wave correlation at t $\approx$ 8.8hrs and of duration $\approx$ 1hr. This is the first observed correlation  for spatially separated gravitational wave detectors.  Two other wave effects (at  $t \approx$ 6.5hrs  in D2 and $t \approx$ 7.3hrs in D1) seen in one detector are masked by the stronger earth-rotation induced minimum in the other detector.  }
  \label{fig:Correlations2}}
\end{figure}

\noindent Fig.\ref{fig:Correlations2} shows two further key tests: 1st the time delay effect in the earth-rotation induced minimum  caused by the detectors not being aligned NS.  The time delay difference  has the value expected. The 2nd effect  is that wave effects are simultaneous, in contrast to the 1st effect.  This is the first coincidence detection of gravitational waves by spatially separated detectors. Soon the separation will be extended to much larger distances.

\begin{figure}
\vspace{-15mm}
\hspace{8mm}\includegraphics[scale=0.45]{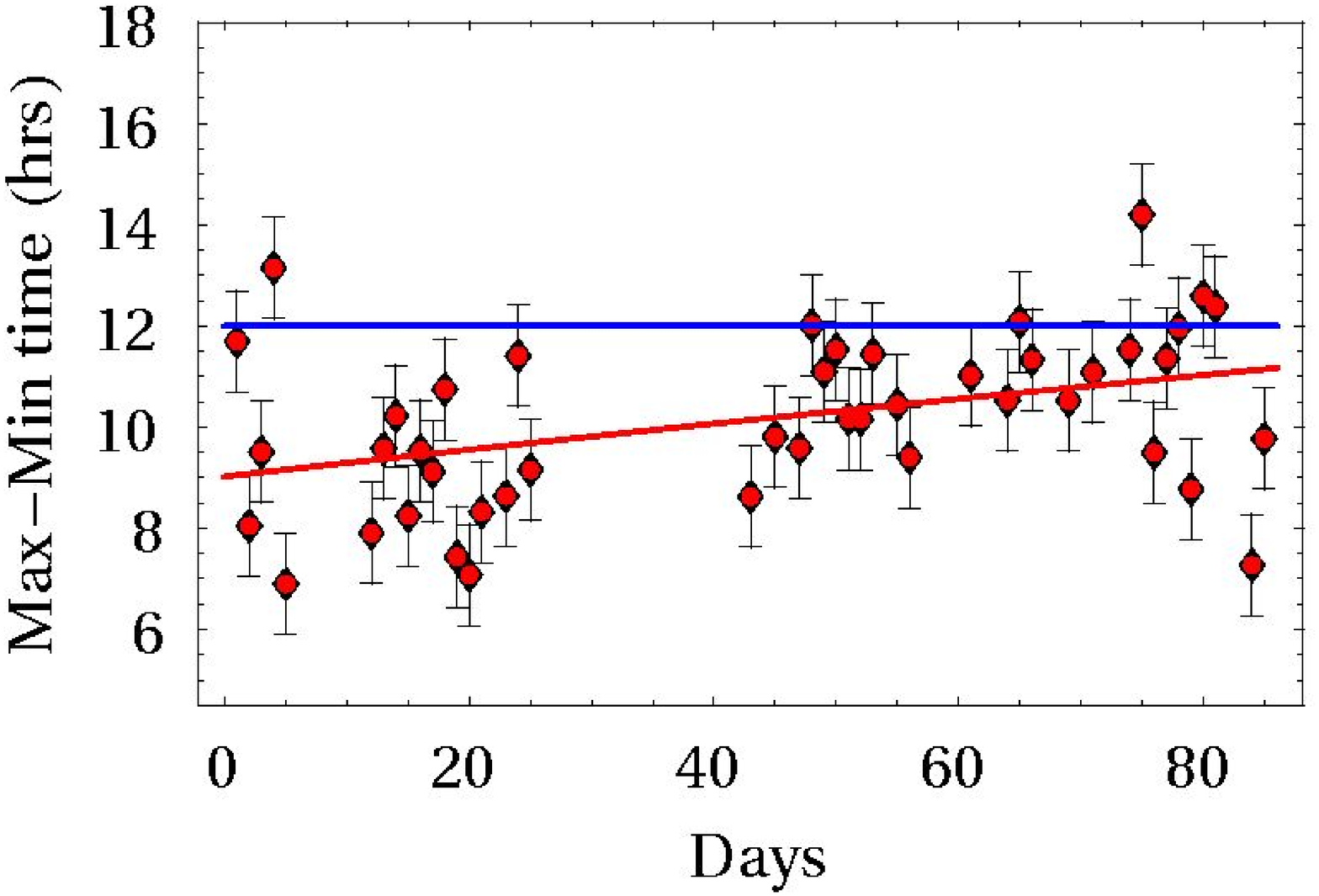}

\hspace{10mm}\includegraphics[scale=0.45]{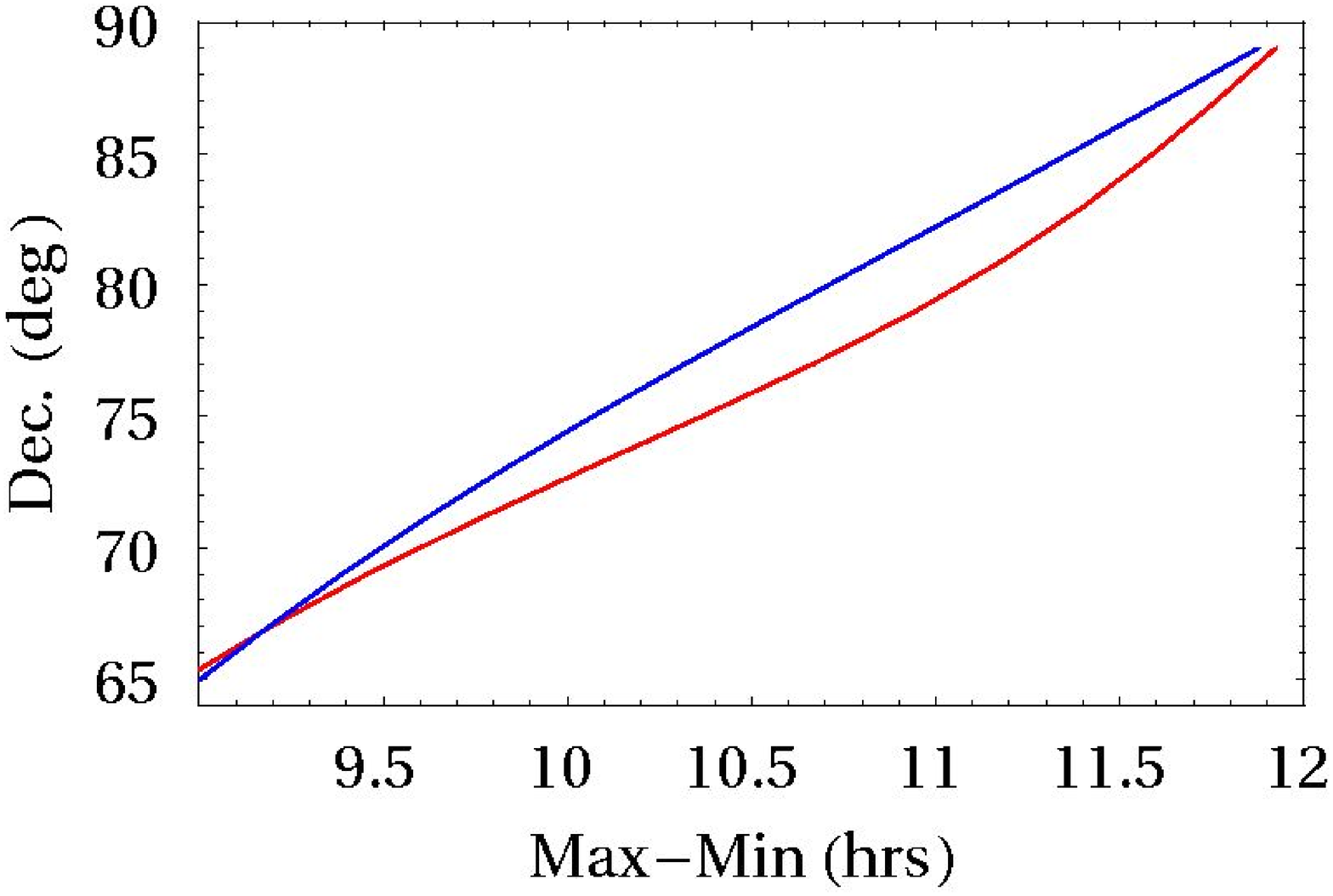}
\vspace{0mm}
\caption{{(a) Time differences between maxima and minima each day from D1, from September 22 to December 16, 2007. Some days are absent due to data logger malfunction. The red curve shows a quadratic best-fit running average.  If the detector arm was orientated along the meridian and there were no wave effects then one would see the horizontal line (blue) at 12hrs.  The data shows, however,  that the running average has a time varying measure, from 9hrs to 11hrs over these days, caused by the orbital motion of the earth about the sun. Wave-effect fluctuations from day to day are also evident.  This data in conjunction with the calibration curve in (b)  permits a determination of the approximate Declination  each day, which is used in the plot shown in Fig.\ref{fig:FT}b. (b) Declination calibration curve for D1 (red) and D2 (blue). From the orientation of the detector, with an offset angle  of $5^0$ for D1 anti-clockwise from the local meridian and $11^0$ for D2 anti-clockwise from the local meridian, and the latitude of Adelaide, the offset angle causes the time duration between a minimum and a maximum to be different from 12hrs, ignoring wave effects.  In conjunction with the running average  in  Fig.\ref{fig:MaxMin}a an approximate determination of the Declination on each day  may be made without needing to also determine the RA and speed.}}
  \label{fig:MaxMin}
\end{figure}

\begin{figure}
\vspace{-27mm}
\hspace{13mm}\includegraphics[scale=0.40]{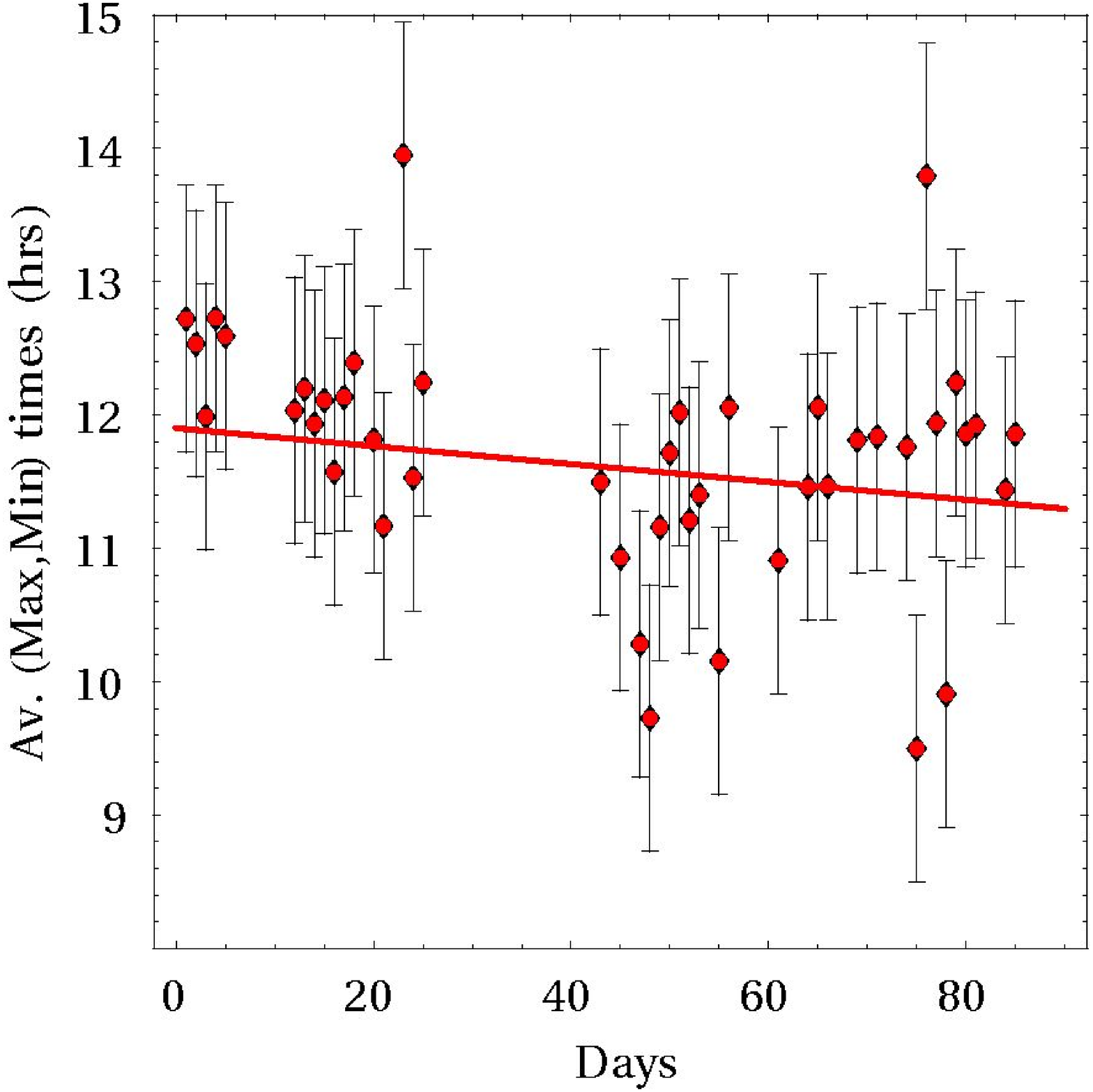}

\hspace{5mm}\includegraphics[scale=0.43]{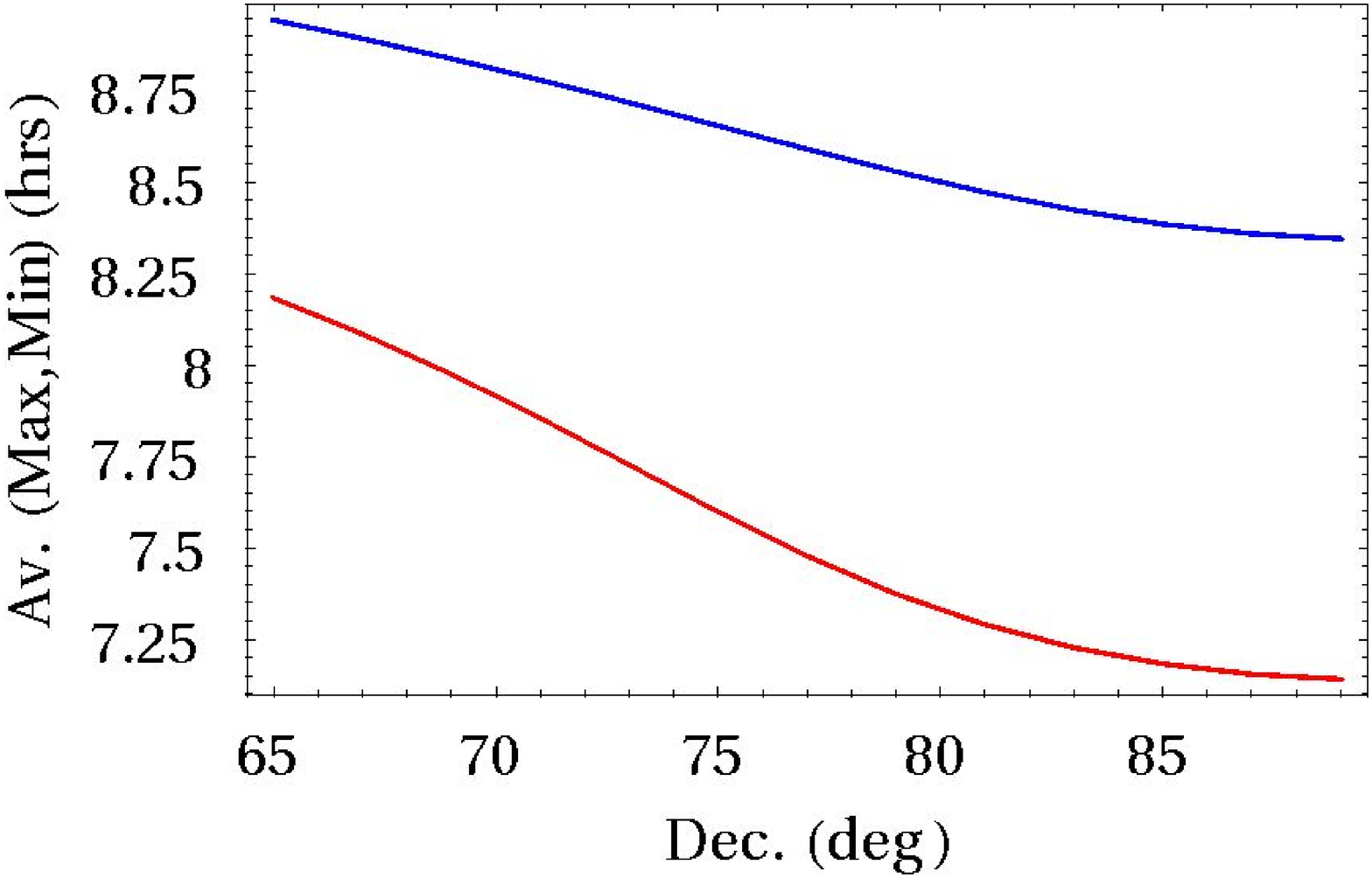}
\vspace{0mm}
\caption{{(a) Time of average of minimum and maximum for each day. The linear best-fit line shows the trend line.  Wave effects are again very evident.  The decreasing trend line is cause by a combination of the sidereal effect,  the earth orbit effect and very low frequency waves.  This data may be used to determine the  approximate RA for each day. However a correction must be applied as the arm offset angle affects the determination.  (b) RA calibration curve for D1 (red) and D2 (blue). The detector  offset angles cause the timing of the mid-point between the minima and maxima, ignoring wave effects, to be delayed in time, beyond  the 6hrs if detector were aligned NS.  So, for example, if the Declination is found to be 70$^0$, then this calibration curve gives  8hrs for D1.  This 8hrs is then subtracted from the time in (a) to give the approximate true local time for the minimum to occur, which then permits computation of the RA for that day.}
  \label{fig:MidMaxMin}}
\end{figure}

\noindent Figs.\ref{fig:MaxMin}  and \ref{fig:MidMaxMin} show the data and calibration curves for the timing of the daily earth-rotation induced minima and maxima over an 80 day period.  Because D1 is orientated  away from the NS these times permit the determination of the Declination (Dec)  and Right Ascension (RA) from the two running averages.  That the running averages change over these 80 days reflects three causes (i) the sidereal time effect, namely that the 3-space velocity vector is related to the positioning of the galaxy, and not the Sun,  (ii) that a smaller component is related to the orbital motion of the earth about the Sun, and (iii) very low frequency wave effects. This analysis gives the changing Dec and RA shown in Fig.\ref{fig:FT}b, giving results which are within  $13^0$ of the 1925/26 Miller results, and for the RA from the DeWitte RF coaxial cable results.  Figs.\ref{fig:MaxMin}a  and \ref{fig:MidMaxMin}a also show the turbulence/wave effects, namely deviations from the running averages.

\noindent Fig.\ref{fig:FT}a shows the frequency analysis of the data. The fourier amplitudes,  which can be related to the strain  $h=v^2/2c^2$, decrease as $f^a$ where the strain spectral index has the value $a=-1.4\pm0.1$, after we allow for the laser noise.

\section{Conclusions}
Sub-mHz gravitational waves have been detected and partially characterised using the optical-fiber
version of a Michelson interferometer. The waves are relatively large and were first detected, though not recognised as such, by Michelson and Morley in 1887.  Since then another 6 experiments \cite{Miller,Torr,DeWitte,CahillCoax,CahillOF1}, including herein, have confirmed the existence of this phenomenon. Significantly three different experimental techniques have been employed, all giving consistent results. In contrast  vacuum-mode Michelson interferometers,  with mechanical mirror support arms, cannot detect this phenomenon due to a design flaw.  A complete characterisation of the waves requires that the optical-fiber detector be calibrated for speed, which means determining the parameter $b$ in (\ref{eqn:intensity}). Then it will be possible to extract the wave component of ${\bf v({\bf r},t)}$ from the average, and so identify the cause of the turbulence/wave effects. A likely candidate is the in-flow of 3-space into the Milky Way central super-massive black hole - this in-flow is responsible for the high,  non-Keplerian, rotation speeds of stars in the galaxy.

The detection of the earth-rotation, earth-orbit and gravitational waves, and over a long period of history, demonstrate that the spacetime formalism of Special Relativity has been very misleading, and that the original Lorentz formalism is the appropriate one; in this the speed of light is not an invariant for all observers, and the Lorentz-Fitzgerald length contraction  and the Lamor time  dilation are real physical effects on rods and clocks in motion through the dynamical 3-space, whereas in the Einstein formalism they are transferred and attributed to a perspective effect of  spacetime, which we now recognise as having no ontological significance - merely a mathematical construct, and in which the invariance of the speed of light is definitional - not observational.

\end{document}